\documentclass[aps,prb,reprint,superscriptaddress,longbibliography]{revtex4-2}
\usepackage{graphicx}
\usepackage{amsmath}
\usepackage{amssymb}
\usepackage{natbib}
\usepackage{hyperref}
\usepackage{xcolor}
\begin{document}
\title{Double-leaf Riemann surface topological converse magnetoelectricity} 
\author{Ying Zhou}
\email{These authors contributed equally to this work}
\author{Haoshen Ye}
\email{These authors contributed equally to this work}
\affiliation{Key Laboratory of Quantum Materials and Devices of Ministry of Education, School of Physics, Southeast University, Nanjing 211189, China}
\author{Junting Zhang}
\email{juntingzhang@cumt.edu.cn}
\affiliation{School of Materials Science and Physics, China University of Mining and Technology, Xuzhou 221116, China}
\author{Shuai Dong}
\email{sdong@seu.edu.cn}
\affiliation{Key Laboratory of Quantum Materials and Devices of Ministry of Education, School of Physics, Southeast University, Nanjing 211189, China}

\begin{abstract}
Electric field control of magnetism in solids, i.e. the converse magnetoelectricity, is highly desired for applications of scalable energy-efficient logic devices. However, it is not only a technical challenge but also a scientific paradox, since in principle the electric and magnetic degrees of freedom obey distinct rules of symmetries. Despite the great progresses obtained in the community of multiferroics during the past decades, the success of magnetoelectricity remains on its way and more alternative approaches with conceptual revolution are urgently needed. Here, by introducing the concept of topology into multiferroics, an exotic magnetoelectric double-leaf Riemann-surface is unveiled based on the mechanism of spin-dependent $d-p$ hybridization in a two-dimensional magnet: GdI$_2$ monolayer. Protected by the topology, a $180^\circ$ spin reversal can be precisely achieved by an electric cycle, leading to a robust and dissipationless converse magnetoelectric function. Such a topological magnetoelectricity allows the nontrivial manipulation of magnetization by AC electric field. In this category, more candidate materials with better performance are designed targetedly, which pave the road to the potential applications with topological magnetoelectrics.
\end{abstract}
\maketitle

\section{Introduction}
\begin{figure}
	\centering
	\includegraphics[width=0.48\textwidth]{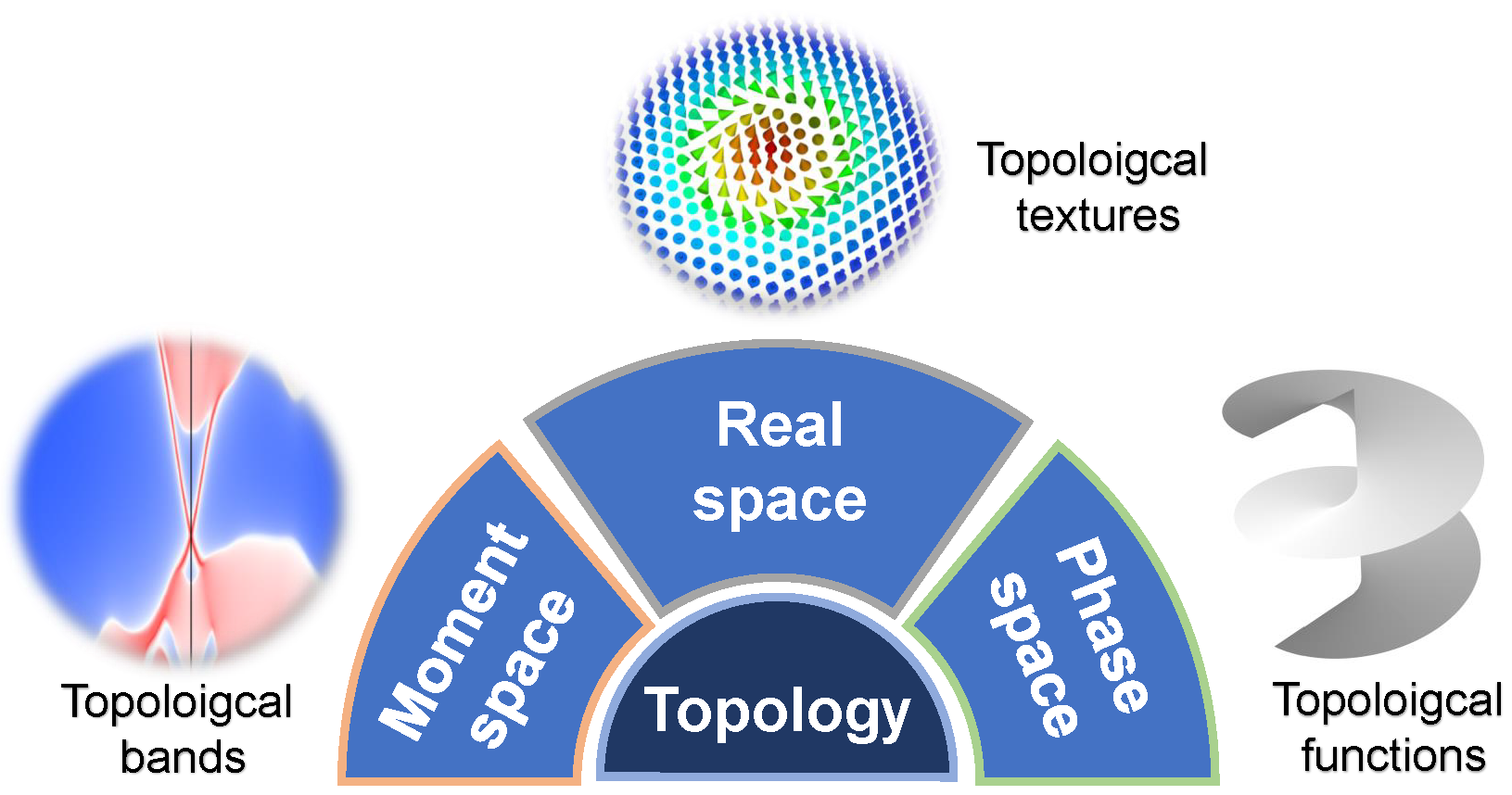}
	\caption{Branches of topology in matter. There are topological band structures in moment space, topological textures of in real space, and topological functions in phase space. These branches are different in physical manifestations, but uniform in their mathematic root.}
	\label{Fig1} 
\end{figure}

Symmetry and topology are two important mathematical concepts, which often govern the physics in explicit or implicit manners. In condensed matter, ideal examples are ferroelectrics and ferromagnets, which break the spatial inversion and time reversal symmetries respectively. And the nontrivial topology can lead to the so-called topological protected physical properties, e.g., dissipationless edge transport in materials with topological electronic/photonic/phononic band structures in moment space \cite{Tschernig:NC,Tang:NP,Kou:PRL}. The concept of topology has also been introduced to the magnetoelectric research. The most intuition cases are some topological ferroic textures, such as vortices/skyrmions/merons of spin moments or electric dipoles \cite{Nagaosa:NN,Fert:NRM,Chauleau:NM,Guo:Science}. These manifestations of topology in moment space and real space form different branches of topological entities in matter, as summarized in Fig.~\ref{Fig1}.

Expected by the demanding of next generation information devices, magnetoelectricity in solids, namely the electric field control of magnetism or magnetic control of electric polarity, is highly valued for applications \cite{Dong:AP,Heron:APR,Wang:PRA}. However, limited by their distinct symmetries, the intrinsic magnetoelectricity are naturally difficult, while other ingredients must be involved as the intermediary to glue the spin polarity and electric polarity \cite{Dong:AP,Dong:NSR}. Till now, it remains challenging to directly and efficiently reverse the magnetization via electrical field, a key magnetoelectric function. Some approaches have been proposed to overcome this difficulty. For example, a two-step $90^\circ$ rotation of magnetization can lead to a net $180^\circ$ reversal, via the piezoelectric tuning of magnetocrystalline anisotropy \cite{Wang:SR}. The electrical tuning of charge disproportion in ferrimagnetic systems can also reverse the net magnetization \cite{Weng:PRL,Lin:PRM}. A recent theory also proposed that the dynamics of a N\'eel-type magnetic domain can resize the ferromagnetic domains during the reversal of chirality \cite{Chen:PRL}.

Very recently, another kind of topology, i.e. topological functions in magnetoelectric phase space (Fig.~\ref{Fig1}), was unveiled in several multiferroics with complex antiferromagnetic structures. In GdMn$_2$O$_5$, a topologically protected magnetoelectric switching coined as ``magnetic crankshaft" was identified, which can convert the increasing-and-decreasing cycle of magnetic field into a circular spin rotation \cite{Ponet:Nature}. While in TbMn$_3$Cr$_4$O$_{12}$, the magnetism induced ferroelectric polarization can span a Roman surface, a three-dimensional non-oriented topological object, which can be traced via an ergodic rotation of spin orientation \cite{Liu:NC,Wang:PRB23}. These intrinsic topologies lead to exotic magnetoelectric functions under magnetic field, which lead to conceptual revolution of magnetoelectricity and extend the territory of topology in condensed matter.

Here we propose another kind of topological magnetoelectric function in phase space, coined as the double-leaf Riemann surface converse magnetoelectricity, which should generally exist in various non-centrosymmetric magnets. For simplicity and practicality, we use the ferromagnetic GdI$_2$ monolayer as the first model system, since its corresponding van der Waals (vdW) bulk is experimentally existing \cite{Mee:IC}. In the following, we will demonstrate how the double-leaf Riemann surface topology provides a robust solution to the long-term challenge of electric field reversal of magnetization. And the general conditions will also be clarified and more candidate materials in this category will also be suggested.

\section{Model \& Methods}
\subsection{First-principles calculations}	
The first-principles calculations based on density functional theory (DFT) were performed using the projector-augmented wave (PAW) method, as implemented in the Vienna \emph{ab initio} Simulation Package (VASP) \cite{Kresse:PRB,Kresse:CMS,BLOCHL:PRB}. The Perdew-Burke-Ernzerhof functional was used as the exchange-correlation functional \cite{Kresse:PRB1,Wang:PRB}. A vacuum space of $20$ {\AA} was added to avoid interaction between neighboring periodic images. We used a cutoff energy of $600$ eV for the plane-wave bases, and a $\varGamma$-centered $15\times15\times1$ $k$-point mesh for the Brillouin zone integration. The in-plane lattice constants and internal atomic coordinates were relaxed until the Hellman-Feynman force on each atom is less than $0.005$ eV/{\AA}. A convergence threshold of $10^{-7}$ eV was used for the electronic self-consistency loop. To describe correlated $4f$ electrons of Gd, the GGA+$U$ method is applied \cite{Dudarev:PRB}, and the adopted $U_{\rm eff}$ is $8$ eV for Gd's $4f$ orbitals \cite{Wang:MH}. For comparison, Heyd-Scuseria-Ernzerhof (HSE06) hybrid functional was also considered \cite{Krukau:JCP}.
	
Phonon band structures were calculated using density functional perturbation theory (DFPT). The phonon frequencies and corresponding eigen-modes were calculated on the basis of the extracted force-constant matrices with a $4\times4\times1$ supercell, as implemented in the PHONOPY code \cite{Togo:SM}. The ferroelectric polarization was calculated by using the Berry phase method \cite{KINGSMITH:PRB}. The unit volume of ferroelectric polarization is measured by atomic layer thickness (removing the thickness of the vacuum layer). In addition, Monte Carlo simulations were performed to verify the magnetic ground states and estimate the magnetic transition temperatures.
	
\subsection{Landu-Lifshitz-Gilbert}
The simulation is based on the classical spin model, which can be written as:
	\begin{equation}
		H=-J_{ij}\sum_{\langle i,j\rangle} {{{\bf{S}}_i} \cdot {{\bf{S}}_j}}  - A\sum\limits_i {{{({\bf{S}}_i^z)}^2}},
		\label{ham}
	\end{equation}
with $J$ and $A$ as the exchange parameters and the anisotropy constant. The nearest neighboring $J_1$, next-nearest neighoring $J_2$ , and third-nearest neighoring $J_3$ are considered, which can be extracted from DFT energies. 
	
In micromagnetic simulation, the spin dynamics are described by the Landu-Lifshitz-Gilbert (LLG) equation:
\begin{equation}
		\frac{{\partial {{\bf S}_i}}}{{\partial t}} =  - \frac{\gamma }{{(1 + {\alpha ^2})}}{\bf S}_i \times ({\bf H}_{eff}^i + \alpha {\bf S}_i \times {\bf H}_{eff}^i),
\end{equation}
where $\alpha=0.01$ is the Gilbert damping constant, $\gamma$ is the Gilbert gyromagnetic ratio, and ${\bf H}_{eff}^i=- \frac{1}{{{M}}}\frac{{\partial H}}{{\partial {\bf S}_i}}$ is the effective field derived from the Hamiltonian $H$ (Eq.~\ref{ham} plus the electrostatic energy $-\sum_i\textbf{E}\cdot\textbf{P}_i$), where $M$ is the magnetic moment of Gd.
	
The LLG equation is solved by the fourth order Runge-Kutta method with time steps $dt \ll T = \frac{{1}}{f}$. The sinusoidal AC electric field is described as ${\bf E}(t) = {\bf E}_0\sin(2\pi ft)$. The pulse sequence used in the four electrode configuration is consisted by alternate half-periodic sinusoidal waves.

\section{Results and Discussion}
\subsection{Static magnetoelectricity in GdI$_2$}
\begin{figure}
	\centering
	\includegraphics[width=0.48\textwidth]{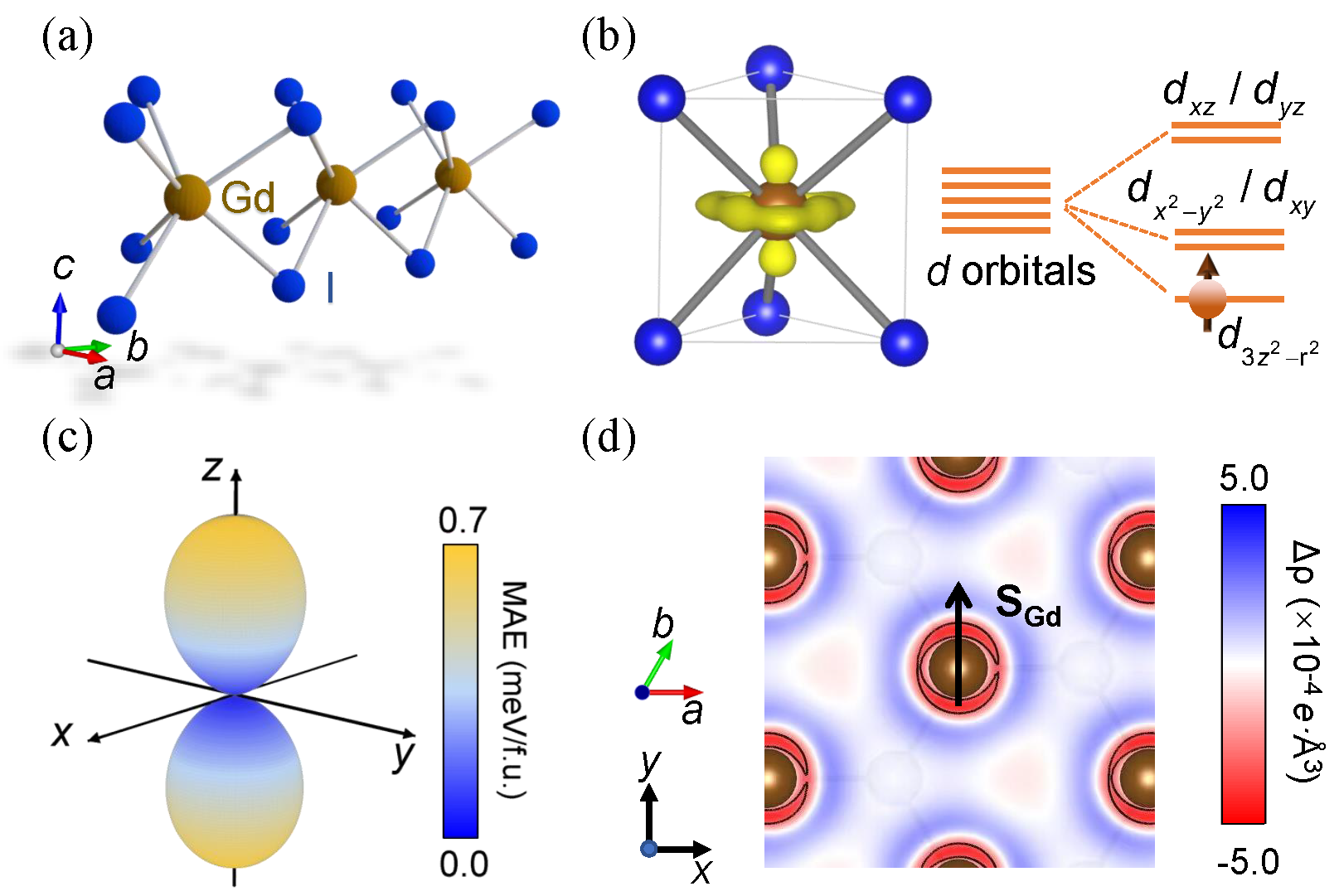}
	\caption{Structure and magnetic properties of GdI$_2$ monolayer. (a) Crystal structure. (b) Schematic of the prism-type crystalline field splitting of GdI$_2$ monolayer. The orbital occupancy of $5d^1$ can be visualized by the electron cloud obtained in DFT calculation. (c) Plot of MAE. The $z$-axis is along the out-of-plane direction. (d) Differential electron density $\Delta\rho(r)$ for Gd's spin along the $y$-axis. Here $\Delta\rho(r)$ is defined as the difference of electron spatial distribution induced by SOC.}
	\label{Fig2} 
\end{figure}

The structure of GdI$_2$ monolayer is non-centrosymmetric and nonpolar (space group $P\overline{6}m2$, No. 187), as shown in Fig.~\ref{Fig2}(a), although its corresponding vdW bulk is centrosymmetric due to its 2H-MoS$_2$-type AB stacking mode \cite{Wang:MH,Mee:IC}. Wang \textit{et al.} proved its possibility of exfoliation \cite{Wang:MH}, with a low cleavage energy $0.26$ J/m$^2$, lower than the experimental value of graphite ($0.36$ J/m$^2$). As shown in Fig.~\ref{Fig1}(b), the Gd$^{2+}$ ion locates at the center of trigonal prism, and there are eight unpaired electrons, forming the $4f^75d^1$ orbital configuration, as confirmed by our density functional theory (DFT) calculation and previous study \cite{Wang:MH}. According to the crystal field of distorted trigonal prism, the residual $5d$ electron stays at a singlet orbital, which is a hybrid one of the $3z^2-r^2$/$x^2-y^2$/$xy$
orbitals \cite{Xiao:Prl}, as visualized in Fig.~\ref{Fig2}(b). The electronic band structure and orbital-projected density of states can be found in Fig.~S1 in Supplementary Materials (SM) \cite{sm}. Such an orbital occupancy leads to an easy-plane magnetocrystalline anisotropy. Our DFT calculation finds the magnetocrystalline anisotropy energy (MAE) is $\sim0.7$ meV/Gd, as shown in Fig.~\ref{Fig1}(c). The local magnetic moment is $8$ $\mu_{\rm B}$/Gd and the magnetic ground state is ferromagnetic (Fig.~S3 and Table~S1 in SM \cite{sm}), as inherited from its vdW bulk with a near room-temperature Curie temperature ($T_{\rm C}=241$ K for monolayer and $313$ K for bulk) \cite{Wang:MH,KASTEN:SSC}. And the monolayer is an insulator with a band gap $0.6$ eV/ $1.1$ eV according to the DFT/ Heyd-Scuseria-Ernzerhof (HSE06) hybrid functional calculation (Fig.~S2 in SM \cite{sm}), in consistent with the previous result \cite{Wang:MH}.

Although its space group $P\overline{6}m2$ is nonpolar, the magnetic point group (MPG) can become polar, depending on the spin orientation. Then a finite ferroelectric polarization can be induced by magnetism, as a character of the type-II multiferroicity \cite{Dong:NSR}. Microscopically, the spin-orbit coupling (SOC) plays a key role to generate the polarization. An intuitive result can be visualized in Fig.~\ref{Fig2}(d): when the spin of Gd$^{2+}$ ($\textbf{S}_{\rm Gd}$) points along the $y$-axis, the contour map of electron density shows slight asymmetry at the Gd site along the $x$-axis, indicating the emergence of electronic polarization along the $x$-axis. The concrete physics is analyzed as follow.

\begin{figure*}
	\centering
	\includegraphics[width=\textwidth]{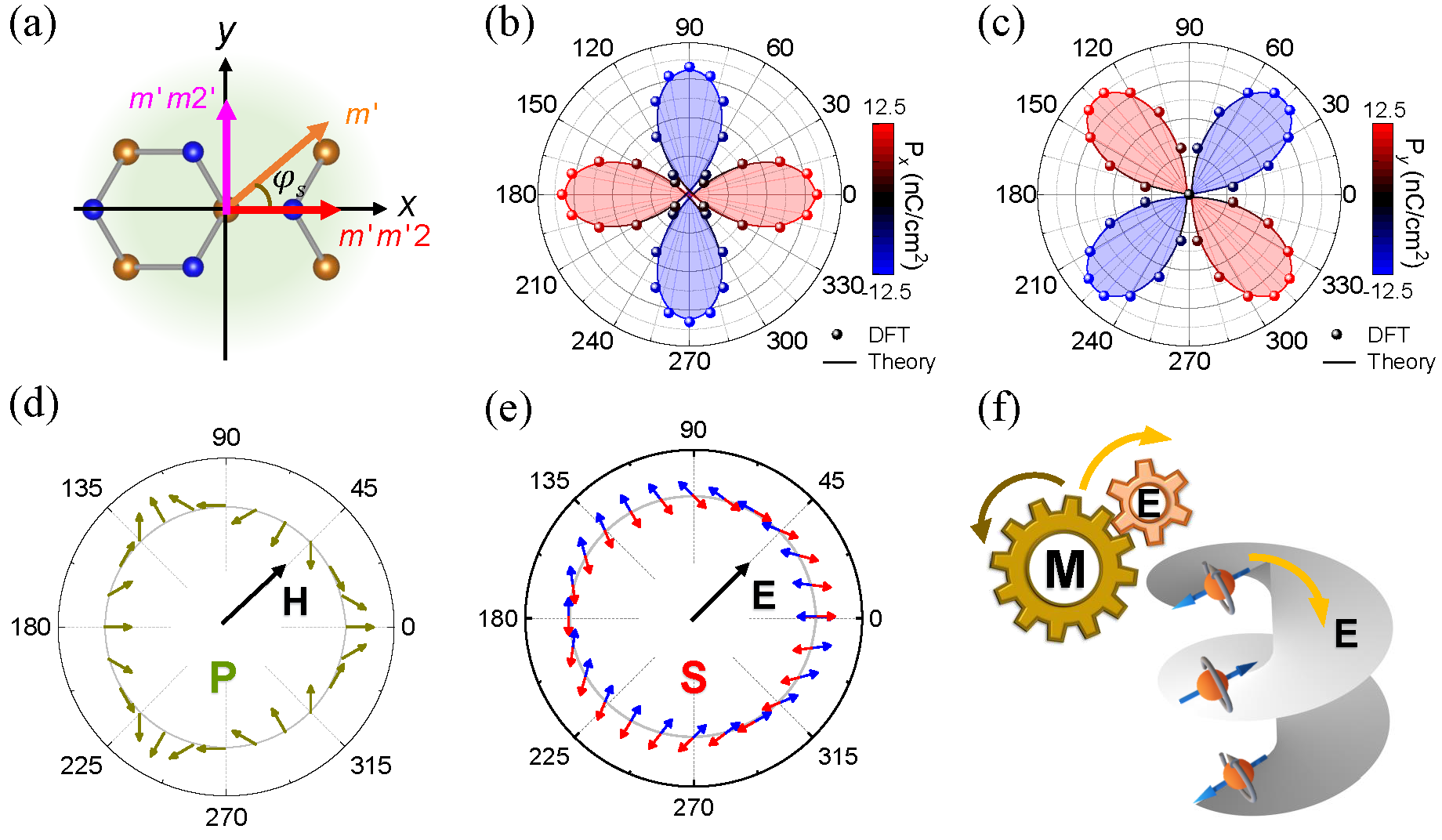}
	\caption{Double-leaf Riemann surface magnetoelectricity in GdI$_2$ monolayer. (a) The evolution of MPG as a function of $\textbf{S}_{\rm Gd}$'s orientation (characterized by its azimuthal angle $\varphi_S$). Orange arrow: generic case of $\textbf{S}_{\rm Gd}$. Red/pink arrows: the special cases of $\textbf{S}_{\rm Gd}$ along the $x$/$y$ axes, which reserve partial symmetries. (b-c) In-plane components of magnetism induced polarization ($P_x$ and $P_y$) as a function of azimuthal angle $\varphi_S$. Dots: numerical results obtained from DFT calculations; Curves: Analytic fitting from the theory of spin-dependent $d-p$ hybridization. (d) The evolution of polarization (green vector) upon the rotation of in-plane magnetic field $\textbf{H}$. (e) The evolution of spin direction (i.e. the magnetization $\textbf{M}$) upon the rotation of in-plane electric field $\textbf{E}$. Red/blue arrows correspond to the odd/even cycle of rotating $\textbf{E}$. (f) The one cycle rotation of $\textbf{E}$ leads to half cycle rotation of $\textbf{M}$, mimicking a gear set with 1:2 perimeters. This magnetoelectric behavior forms the topologic double-leaf Riemann surface, which is a non-oriented surface with a half twist. Then protected by the topology, only even winding number of $\textbf{E}$ can restore $\textbf{M}$, while any odd winding number of $\textbf{E}$ can only flip $\textbf{M}$ by $180^\circ$ robustly and precisely.}
	\label{Fig3}
\end{figure*}

With the ferromagnetic $\textbf{S}_{\rm Gd}$ lying in the $xy$ plane, the out-of-plane $3$-fold rotation ($C_3$) symmetry is broken, which allows the in-plane polarization. For example, when $\textbf{S}_{\rm Gd}$ is along the $x$-axis which is a 2-fold rotation axis, the MPG becomes $m'm'2$, whose polar axis is also along the $x$-axis. The MPG becomes $m'm2'$ once $\textbf{S}_{\rm Gd}$ pointing along the $y$-axis, but its polar axis remains along the $x$-axis. For $\textbf{S}_{\rm Gd}$ along other arbitrary in-plane directions, the MPG becomes $m'$, which allows both nonzero $x$- and $y$-components of polarization (i.e. $P_x$ and $P_y$), as shown in Fig.~\ref{Fig3}(a) and Table~S2 in SM \cite{sm}.

Above symmetry analysis has given a phenomenological and qualitative description of magnetoelectricity, and its microscopic quantum mechanism will be clarified as follow. Due to the ferromagnetic configuration, neither the inverse Dzyaloshinskii-Moriya interaction nor the exchange striction is responsibale for the magnetism induced polarization. Then the possible route to type-II multiferroicity here is the so-called spin-dependent $p-d$ hybridization mechanism, which originates from the modification $p-d$ hybridization by SOC \cite{Murakawa:PRL,Zhang:PRL,Arima:JPSJ}. Analytically, the magnetism induced polarization can be obtained by summing the charge transfers between the magnetic ion and its six ligands \cite{Jia:PRB}: 
\begin{equation}
	\textbf{P}\propto\sum_{i\in[1-6]}(\textbf{S}_{\rm Gd}\cdot\textbf{e}_i)^2\cdot\textbf{e}_i,
	\label{pd}
\end{equation} 
where $\textbf{e}_i$ is the unit vector of the bond between the magnetic ion and ligand. Once $\textbf{S}_{\rm Gd}$ lies in the $xy$ plane, it is straightforward to obtain the analytic expression: 
\begin{equation}
	\begin{array}{*{20}{l}}
		\textbf{P}&=(P_x, P_y, P_z)\\
		&\propto(S_x^2 - S_y^2, -2S_xS_y, 0)\\
		&=[\cos(2\varphi_{\rm{\textit{S}}}), -\sin(2\varphi _{\rm{\textit{S}}}), 0]
	\end{array},
\end{equation} 
where $\varphi_S$ is the azimuthal angle of $\textbf{S}_{\rm Gd}$ as defined in Fig.~\ref{Fig3}(a). The polar plots of $P_x$ and $P_y$ as a function of $\varphi_S$ are shown as curves in Figs.~\ref{Fig3}(b) and ~\ref{Fig3}(c), respectively.

Such analytic results are further confirmed by our DFT calculations [dots in Figs.~\ref{Fig3}(b-c)], implying that the spin-dependent $p-d$ hybridization is indeed the underlying mechanism. The magnitude of $\textbf{P}$ is a constant, reaching $125$ $\mu$C/m$^2$ (by using the thickness of GdI$_2$ monolayer to estimate the volume), which is comparable to other multiferroics in this category (e.g. $100$ $\mu$C/m$^2$ for Ba$_2$CoGe$_2$O$_7$ and $21.2$ $\mu$C/m$^2$ for Ca$_3$FeReO$_7$)\cite{Murakawa:PRL,Zhang:PRL}.

Considering the direct coupling between spin orientation (i.e. the magnetization $\textbf{M}$) and polarization $\textbf{P}$, it is natural to expect intrinsic magnetoelectricity, namely using the magnetic field $\textbf{H}$ to control $\textbf{P}$ (i.e. the direct magnetoelectric effect, DME), or using the electric field $\textbf{E}$ to control $\textbf{M}$ (i.e. CME). As shown in Fig.~\ref{Fig3}(d), by rotating the in-plane magnetic field $\textbf{H}$ quasi-statically for a cycle, the magnetization (i.e. $\textbf{S}_{\rm Gd}$) will rotate synchronously, but the induced polarization $\textbf{P}$ will rotate with a twice angular velocity. Conversely, as shown in Fig.~\ref{Fig3}(e), by rotating the in-plane electric field $\textbf{E}$ quasi-statically for a cycle, the polarization $\textbf{P}$ will rotate synchronously, but the associated $\textbf{S}_{\rm Gd}$ will rotate with a half angular velocity. Thus, a cycle of quasi-static $\textbf{E}$ will reverse $\textbf{M}$ for $180^\circ$, which is a long-desired function of CME. Then two continuous cycles of quasi-static $\textbf{E}$ will restore $\textbf{M}$ to its original direction.

The magnetoelectricity of GdI$_2$ monolayers can be summarized in Fig.~\ref{Fig3}(f). The coupling between electric and magnetic degrees of freedom can be analogous to a gear set with 1:2 perimeters. Mathematically, such a magnetoelectric behavior obeys the topoloy of double-leaf Riemann surface. Only even winding number of $\textbf{E}$ can restore $\textbf{M}$, while any odd winding number of $\textbf{E}$ can only flip $\textbf{M}$ by $180^\circ$, protected by the topology of double-leaf Riemann surface which originates from the mathematic formula of $(\textbf{S}\cdot\textbf{e})^2$. In addition, since both the $|\textbf{P}|$ and MAE are isotropic in the $xy$ plane, ideally such an in-plane rotation of quasi-static $\textbf{E}$ or $\textbf{H}$ does not need to overcome any energy barriers, and thus dissipationless.

\subsection{Dynamic magnetoelectricity in GdI$_2$}

\begin{figure*}
	\centering
	\includegraphics*[width=\textwidth]{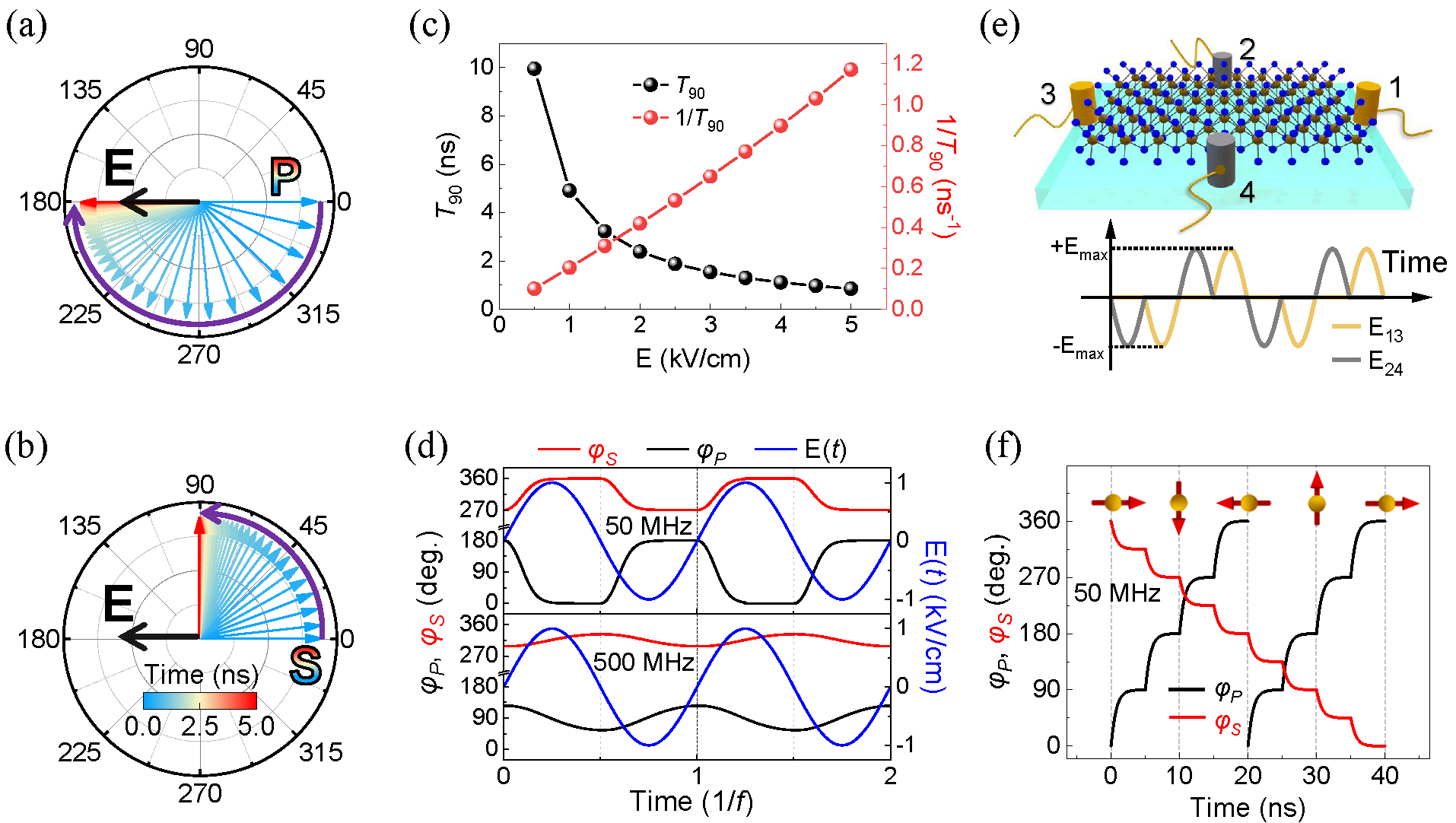}
	\caption{Electrical response of $\textbf{P}$ and $\textbf{S}_{\rm Gd}$. The rotation dynamics of $\textbf{P}$ (a) and $\textbf{S}_{\rm Gd}$ (b) excited by static electric field of $1$ kV/cm. The black arrow denotes direction of electric field. (c) The evolution of switching duration $T_{90}$ (and its reciprocal $1/T_{90}$ ) as a function of static electric field $\textbf{E}$. (d) Oscillation of $\textbf{P}$ and $\textbf{S}_{\rm Gd}$ excited by sinusoidal AC electric field E($t$) of $f$ = $50$ MHz (top) and $f$ = $500$ MHz (bottom). $\varphi_S$ and $\varphi_P$ are azimuthal angles of $\textbf{S}_{\rm Gd}$ and the corresponding polarization $\textbf{P}$. (e) Schematic diagram of four electrodes in perpendicular configuration. Inset: a period of alternate electric field pulses. (f) The half-harmonic dynamics of $\textbf{S}_{\rm Gd}$ excited by pulse-width modulated electric field sets of $f$ = $50$ MHz. Inset: direction of $\textbf{S}_{\rm Gd}$.  In all cases, initial $\textbf{P}$ and $\textbf{S}_{\rm Gd}$ are oriented to the +$x$ direction, and the maximum of AC electric fields are $1$ kV/cm.}
	\label{Fig4}
\end{figure*}

Above topological double-leaf Riemann surface magnetoelectricity is straightforward based on the protocol of quasi-static electric/magnetic cycles. However, for information reading/writing in potential magnetoelectric logic devices, dynamic process must be involved. Here we employ the Landau-Lifshitz-Gilbert (LLG) equation to describe the dynamics of $\textbf{S}_{\rm Gd}$ and associated polarization.

First, starting from the case with $\textbf{S}_{\rm Gd}$ along the $x$-axis (i.e. $\textbf{P}$ also along the $x$-axis), the intrinsic switching duration is estimated by applying a static electric field $\textbf{E}$ along the $-x$ direction. As shown in Figs.~\ref{Fig4}(a-b) and movie S1 in SM \cite{sm}, under a moderate field $1$ kV/cm, $\textbf{P}$ is reversed within $5$ ns, accompanying by a $90^\circ$ rotation of $\textbf{S}_{\rm Gd}$. This dynamic process is fast in the beginning and becomes slower in the home stretch. This switching duration ($T_{90}$) is in proportional to the $1/|\textbf{E}\cdot\textbf{P}|$, as expected from the LLG equation [Fig.~\ref{Fig4}(c) and Fig. ~S4 in SM \cite{sm}]. Thus, the larger $\textbf{E}$ (or $\textbf{P}$), the faster magnetoelectric switching.

Second, once an AC electric field $\textbf{E}(t)$ is applied along the $x$-axis, the dynamic behavior becomes frequency dependent, as shown in Fig.~\ref{Fig4}(d) and movie S2 in SM \cite{sm}. In the low-frequency region, e.g. $f=50$ MHz, the dynamics of spin and dipole can fully follow the AC $\textbf{E}(t)$, forth and back periodically [Fig.~\ref{Fig4}(d) and Fig. ~S5(a) in SM \cite{sm}]. In the high frequency cases, e.g. $f=500$ MHz, $\textbf{S}_{\rm Gd}$ and $\textbf{P}$ can not follow $\textbf{E}(t)$ anymore, the flippings of spin and dipole are only partially and oscillating around the middle point: $90^\circ$ (or $270^\circ$) for $\textbf{P}$ and $315^\circ$ (or $45^\circ$) for $\textbf{S}_{\rm Gd}$ [Fig.~\ref{Fig4}(d) and Fig. ~S5(b) in SM \cite{sm}].

Third, to obtain the topological winding of $\textbf{P}$ and $\textbf{S}_{\rm Gd}$, four electrodes are arranged in the perpendicular configuration, as depicted in Fig.~\ref{Fig4}(e). Then a sequence of alternate electric field pulses ($E_{13}$ \& $E_{24}$) is applied, also shown in Fig.~\ref{Fig4}(e) and Fig. ~S6 in SM \cite{sm}. At low frequency, e.g. $f=50$ MHz, each pulse can generate a $90^\circ$ rotation of $\textbf{P}$ and $45^\circ$ rotation of $\textbf{S}_{\rm Gd}$. Then eight pulses leads to a full winding cycle of $\textbf{S}_{\rm Gd}$ and twice cycles of $\textbf{P}$, as shown in Fig.~\ref{Fig4}(f) and movie S3 in SM \cite{sm}.

Finally, it should be noted that the surrounding magnetic field generated by the AC electric field orientates to the out-of-plane direction. Its Zeeman energy is in the order of $\sim0.15$ $\mu$eV/Gd, which is negligible compared to its MAE ($0.7$ meV/Gd). Thus, this double-leaf Riemann surface magnetoelectricity with in-plane rotations of $\textbf{S}_{\rm Gd}$ and $\textbf{P}$ is electrically driven.

\subsection{General rules \& more candidates}

\begin{figure}
	\centering
	\includegraphics[width=0.48\textwidth]{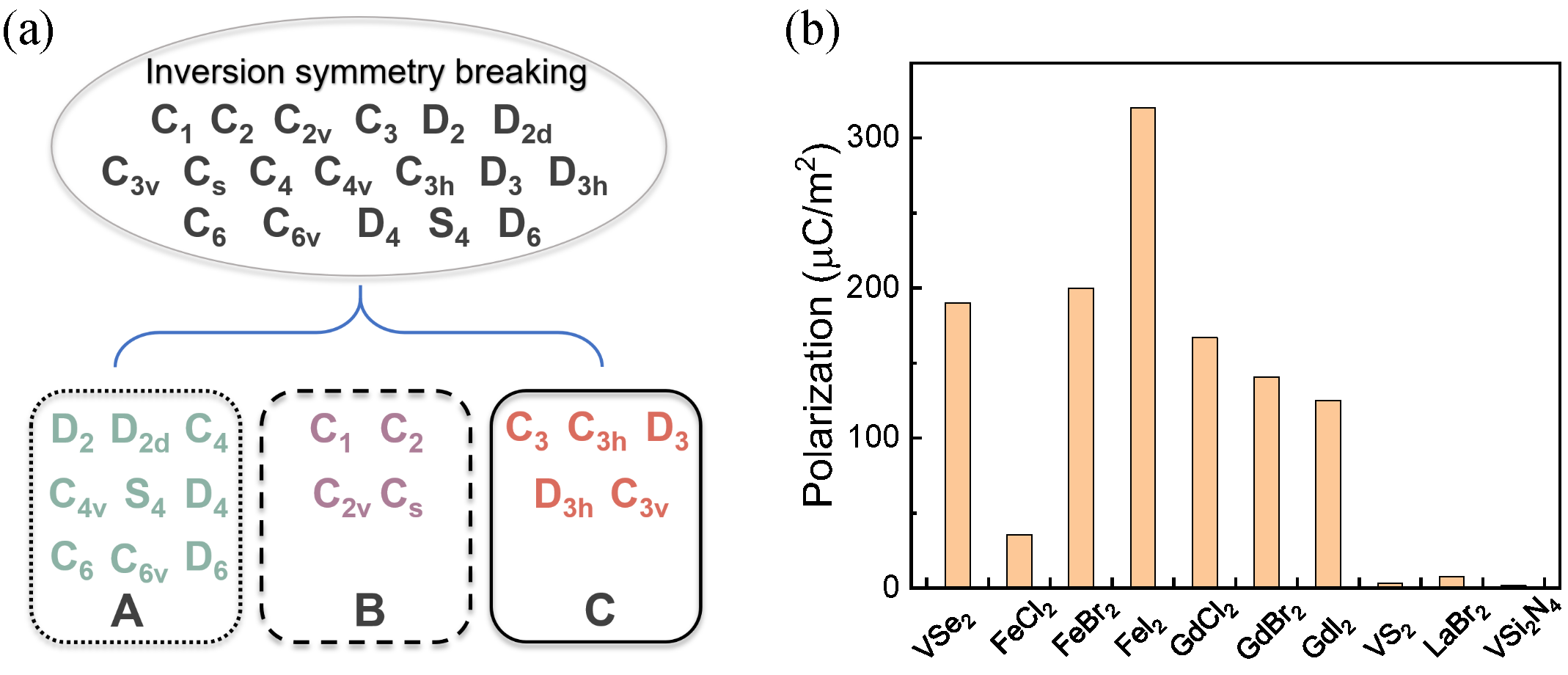}
	\caption{Symmetry classification and more candidates. (a) Symmetry classification of 2D noncentrosymmetric point groups in terms of spin-induced polarization. The ferromagnetic state with in-plane spins are the preconditions. Class A: the $p$-$d$ hybridization mechanism can not an in-plane polarization. Class B: double-leaf Riemann converse magnetoelectricity is possible but not guaranteed. And the dissipation of converse magnetoelectricity is unavoidable. Class C: double-leaf Riemann converse magnetoelectricity is guaranteed and dissipationless. (b) The magnitude of spin-induced polarization of some selected 2D ferromagnets in the $D_{3h}$ category.}
	\label{Fig5} 
\end{figure}

Although above studies only focused on GdI$_2$ monolayer, the physical mechanism should generally apply to more systems. The spin-dependent $p$-$d$ hybridization can induce ferroelectricity only in those point groups lacking inversion symmetry. In the 2D case, there are eighteen noncentrosymmetric point groups. Their magnetoelectric tensors are derived, as summarized in Supplementary text. According to these tensors, half of them [class A in Fig.~\ref{Fig5}(a)] do not permit the in-plane polarization originated from spin-dependent $p$-$d$ hybridization when the ferromagnetic spins lying in the $ab$-plane.

For those point groups in class B, although there is still a $1:2$ relationship between the spin and polarization modulation period, the double-leaf Riemann converse magnetoelectricity is not guaranteed, since the trajectory of polarization may not pass through all four quadrants (e.g. Fig.~S7(a) and Fig.~S8(b) in SM \cite{sm}). Even if the trajectory of polarization can pass through all four quadrants (e.g. Fig.~S7(b-c) and Fig.~S8(a) in SM \cite{sm}), the amplitude of polarization is not a constant, but depends on its orientation. Further, the in-plane magnetocrystalline anisotropy will arise. These two factors establish energy barriers during the spin/dipole rotation, which lead to energy dissipation for converse magnetoelectricity.

Only those point groups in class C, which contain a common 3-fold rotational symmetry ($C_3$), allow the dissipationless converse magnetoelectricity (e.g. Fig.~S9 in SM \cite{sm}). In addition to these symmetry requirements, the candidate systems should be ferromagnetic and the $ab$-plane should be the magnetic easy plane.

Following above conditions, several materials were screened for this magnetoelectricity, including the $2H$-$MX_2$ type  and VSi$_2$N$_4$ monolayer. The $p$-$d$ hybridization mechanism can indeed induce ferroelectric polarization in these systems, while their amplitude depends on their electronic structures especially their SOC strength, as shown in Fig.~\ref{Fig5}(b). Despite their values of polarization, all of them exhibit the double-leaf Riemann magnetoelectricity and dissipationless converse magnetoelectric function.

\section{Conclusion}
In summary, a double-leaf Riemann surface topological magnetoelectricity has been predicted, in which the topology protects the $180^\circ$ reversal of magnetization under a cycle of electric field. Based on symmetry analysis, $p-d$ hybridization theory, and DFT calculations, the double-leaf Riemann surface topological magnetoelectricity has been demonstrated in GdI$_2$ monolayer and other candidates. Such a topological magnetoelectricity allows the direct and precise manipulation of magnetization by AC electric fields. Our work extends the territory of topology in condensed matter, and provides a robust solution to the long-term challenge of electrical reversal of magnetism.

\begin{acknowledgments}
We thank Profs. Yunfeng Jiang and Yisheng Chai for helpful disucssions on the mathematical formula of magnetoelectricity. This work was supported by the National Natural Science Foundation of China (Grant Nos. 12325401, 12274069, \& 12374097) and the Big Data Computing Center of Southeast University.
\end{acknowledgments}

\bibliography{reference}

\begin{thebibliography}{37}%
\makeatletter
\providecommand \@ifxundefined [1]{%
 \@ifx{#1\undefined}
}%
\providecommand \@ifnum [1]{%
 \ifnum #1\expandafter \@firstoftwo
 \else \expandafter \@secondoftwo
 \fi
}%
\providecommand \@ifx [1]{%
 \ifx #1\expandafter \@firstoftwo
 \else \expandafter \@secondoftwo
 \fi
}%
\providecommand \natexlab [1]{#1}%
\providecommand \enquote  [1]{``#1''}%
\providecommand \bibnamefont  [1]{#1}%
\providecommand \bibfnamefont [1]{#1}%
\providecommand \citenamefont [1]{#1}%
\providecommand \href@noop [0]{\@secondoftwo}%
\providecommand \href [0]{\begingroup \@sanitize@url \@href}%
\providecommand \@href[1]{\@@startlink{#1}\@@href}%
\providecommand \@@href[1]{\endgroup#1\@@endlink}%
\providecommand \@sanitize@url [0]{\catcode `\\12\catcode `\$12\catcode
  `\&12\catcode `\#12\catcode `\^12\catcode `\_12\catcode `\%12\relax}%
\providecommand \@@startlink[1]{}%
\providecommand \@@endlink[0]{}%
\providecommand \url  [0]{\begingroup\@sanitize@url \@url }%
\providecommand \@url [1]{\endgroup\@href {#1}{\urlprefix }}%
\providecommand \urlprefix  [0]{URL }%
\providecommand \Eprint [0]{\href }%
\providecommand \doibase [0]{https://doi.org/}%
\providecommand \selectlanguage [0]{\@gobble}%
\providecommand \bibinfo  [0]{\@secondoftwo}%
\providecommand \bibfield  [0]{\@secondoftwo}%
\providecommand \translation [1]{[#1]}%
\providecommand \BibitemOpen [0]{}%
\providecommand \bibitemStop [0]{}%
\providecommand \bibitemNoStop [0]{.\EOS\space}%
\providecommand \EOS [0]{\spacefactor3000\relax}%
\providecommand \BibitemShut  [1]{\csname bibitem#1\endcsname}%
\let\auto@bib@innerbib\@empty
\bibitem [{\citenamefont {Tschernig}\ \emph {et~al.}(2021)\citenamefont
  {Tschernig}, \citenamefont {Jimenez-Galan}, \citenamefont {Christodoulides},
  \citenamefont {Ivanov}, \citenamefont {Busch}, \citenamefont {Bandres},\ and\
  \citenamefont {Perez-Leija}}]{Tschernig:NC}%
  \BibitemOpen
  \bibfield  {author} {\bibinfo {author} {\bibfnamefont {K.}~\bibnamefont
  {Tschernig}}, \bibinfo {author} {\bibfnamefont {A.}~\bibnamefont
  {Jimenez-Galan}}, \bibinfo {author} {\bibfnamefont {D.~N.}\ \bibnamefont
  {Christodoulides}}, \bibinfo {author} {\bibfnamefont {M.}~\bibnamefont
  {Ivanov}}, \bibinfo {author} {\bibfnamefont {K.}~\bibnamefont {Busch}},
  \bibinfo {author} {\bibfnamefont {M.~A.}\ \bibnamefont {Bandres}},\ and\
  \bibinfo {author} {\bibfnamefont {A.}~\bibnamefont {Perez-Leija}},\
  }\bibfield  {title} {\bibinfo {title} {Topological protection versus degree
  of entanglement of two-photon light in photonic topological insulators},\
  }\href@noop {} {\bibfield  {journal} {\bibinfo  {journal} {Nat. Commun.}\
  }\textbf {\bibinfo {volume} {12}},\ \bibinfo {pages} {1974} (\bibinfo {year}
  {2021})}\BibitemShut {NoStop}%
\bibitem [{\citenamefont {Tang}\ \emph {et~al.}(2017)\citenamefont {Tang},
  \citenamefont {Zhang}, \citenamefont {Wong}, \citenamefont {Pedramrazi},
  \citenamefont {Tsai}, \citenamefont {Jia}, \citenamefont {Moritz},
  \citenamefont {Claassen}, \citenamefont {Ryu}, \citenamefont {Kahn},
  \citenamefont {Jiang}, \citenamefont {Yan}, \citenamefont {Hashimoto},
  \citenamefont {Lu}, \citenamefont {Moore}, \citenamefont {Hwang},
  \citenamefont {Hwang}, \citenamefont {Hussain}, \citenamefont {Chen},
  \citenamefont {Ugeda}, \citenamefont {Liu}, \citenamefont {Xie},
  \citenamefont {Devereaux}, \citenamefont {Crommie}, \citenamefont {Mo},\ and\
  \citenamefont {Shen}}]{Tang:NP}%
  \BibitemOpen
  \bibfield  {author} {\bibinfo {author} {\bibfnamefont {S.}~\bibnamefont
  {Tang}}, \bibinfo {author} {\bibfnamefont {C.}~\bibnamefont {Zhang}},
  \bibinfo {author} {\bibfnamefont {D.}~\bibnamefont {Wong}}, \bibinfo {author}
  {\bibfnamefont {Z.}~\bibnamefont {Pedramrazi}}, \bibinfo {author}
  {\bibfnamefont {H.-Z.}\ \bibnamefont {Tsai}}, \bibinfo {author}
  {\bibfnamefont {C.}~\bibnamefont {Jia}}, \bibinfo {author} {\bibfnamefont
  {B.}~\bibnamefont {Moritz}}, \bibinfo {author} {\bibfnamefont
  {M.}~\bibnamefont {Claassen}}, \bibinfo {author} {\bibfnamefont
  {H.}~\bibnamefont {Ryu}}, \bibinfo {author} {\bibfnamefont {S.}~\bibnamefont
  {Kahn}}, \bibinfo {author} {\bibfnamefont {J.}~\bibnamefont {Jiang}},
  \bibinfo {author} {\bibfnamefont {H.}~\bibnamefont {Yan}}, \bibinfo {author}
  {\bibfnamefont {M.}~\bibnamefont {Hashimoto}}, \bibinfo {author}
  {\bibfnamefont {D.}~\bibnamefont {Lu}}, \bibinfo {author} {\bibfnamefont
  {R.~G.}\ \bibnamefont {Moore}}, \bibinfo {author} {\bibfnamefont {C.-C.}\
  \bibnamefont {Hwang}}, \bibinfo {author} {\bibfnamefont {C.}~\bibnamefont
  {Hwang}}, \bibinfo {author} {\bibfnamefont {Z.}~\bibnamefont {Hussain}},
  \bibinfo {author} {\bibfnamefont {Y.}~\bibnamefont {Chen}}, \bibinfo {author}
  {\bibfnamefont {M.~M.}\ \bibnamefont {Ugeda}}, \bibinfo {author}
  {\bibfnamefont {Z.}~\bibnamefont {Liu}}, \bibinfo {author} {\bibfnamefont
  {X.}~\bibnamefont {Xie}}, \bibinfo {author} {\bibfnamefont {T.~P.}\
  \bibnamefont {Devereaux}}, \bibinfo {author} {\bibfnamefont {M.~F.}\
  \bibnamefont {Crommie}}, \bibinfo {author} {\bibfnamefont {S.-K.}\
  \bibnamefont {Mo}},\ and\ \bibinfo {author} {\bibfnamefont {Z.-X.}\
  \bibnamefont {Shen}},\ }\bibfield  {title} {\bibinfo {title} {Quantum spin
  hall state in monolayer 1{T$'$-WT}e$_2$},\ }\href@noop {} {\bibfield
  {journal} {\bibinfo  {journal} {Nat. Phys.}\ }\textbf {\bibinfo {volume}
  {13}},\ \bibinfo {pages} {683} (\bibinfo {year} {2017})}\BibitemShut
  {NoStop}%
\bibitem [{\citenamefont {Kou}\ \emph {et~al.}(2014)\citenamefont {Kou},
  \citenamefont {Guo}, \citenamefont {Fan}, \citenamefont {Pan}, \citenamefont
  {Lang}, \citenamefont {Jiang}, \citenamefont {Shao}, \citenamefont {Nie},
  \citenamefont {Murata}, \citenamefont {Tang}, \citenamefont {Wang},
  \citenamefont {He}, \citenamefont {Lee}, \citenamefont {Lee},\ and\
  \citenamefont {Wang}}]{Kou:PRL}%
  \BibitemOpen
  \bibfield  {author} {\bibinfo {author} {\bibfnamefont {X.}~\bibnamefont
  {Kou}}, \bibinfo {author} {\bibfnamefont {S.-T.}\ \bibnamefont {Guo}},
  \bibinfo {author} {\bibfnamefont {Y.}~\bibnamefont {Fan}}, \bibinfo {author}
  {\bibfnamefont {L.}~\bibnamefont {Pan}}, \bibinfo {author} {\bibfnamefont
  {M.}~\bibnamefont {Lang}}, \bibinfo {author} {\bibfnamefont {Y.}~\bibnamefont
  {Jiang}}, \bibinfo {author} {\bibfnamefont {Q.}~\bibnamefont {Shao}},
  \bibinfo {author} {\bibfnamefont {T.}~\bibnamefont {Nie}}, \bibinfo {author}
  {\bibfnamefont {K.}~\bibnamefont {Murata}}, \bibinfo {author} {\bibfnamefont
  {J.}~\bibnamefont {Tang}}, \bibinfo {author} {\bibfnamefont {Y.}~\bibnamefont
  {Wang}}, \bibinfo {author} {\bibfnamefont {L.}~\bibnamefont {He}}, \bibinfo
  {author} {\bibfnamefont {T.-K.}\ \bibnamefont {Lee}}, \bibinfo {author}
  {\bibfnamefont {W.-L.}\ \bibnamefont {Lee}},\ and\ \bibinfo {author}
  {\bibfnamefont {K.~L.}\ \bibnamefont {Wang}},\ }\bibfield  {title} {\bibinfo
  {title} {Scale-invariant quantum anomalous hall effect in magnetic
  topological insulators beyond the two-dimensional limit},\ }\href@noop {}
  {\bibfield  {journal} {\bibinfo  {journal} {Phys. Rev. Lett.}\ }\textbf
  {\bibinfo {volume} {113}},\ \bibinfo {pages} {137201} (\bibinfo {year}
  {2014})}\BibitemShut {NoStop}%
\bibitem [{\citenamefont {Nagaosa}\ and\ \citenamefont
  {Tokura}(2013)}]{Nagaosa:NN}%
  \BibitemOpen
  \bibfield  {author} {\bibinfo {author} {\bibfnamefont {N.}~\bibnamefont
  {Nagaosa}}\ and\ \bibinfo {author} {\bibfnamefont {Y.}~\bibnamefont
  {Tokura}},\ }\bibfield  {title} {\bibinfo {title} {Topological properties and
  dynamics of magnetic skyrmions},\ }\href@noop {} {\bibfield  {journal}
  {\bibinfo  {journal} {Nat. Nanotechnol}\ }\textbf {\bibinfo {volume} {8}},\
  \bibinfo {pages} {899} (\bibinfo {year} {2013})}\BibitemShut {NoStop}%
\bibitem [{\citenamefont {Fert}\ \emph {et~al.}(2017)\citenamefont {Fert},
  \citenamefont {Reyren},\ and\ \citenamefont {Cros}}]{Fert:NRM}%
  \BibitemOpen
  \bibfield  {author} {\bibinfo {author} {\bibfnamefont {A.}~\bibnamefont
  {Fert}}, \bibinfo {author} {\bibfnamefont {N.}~\bibnamefont {Reyren}},\ and\
  \bibinfo {author} {\bibfnamefont {V.}~\bibnamefont {Cros}},\ }\bibfield
  {title} {\bibinfo {title} {Magnetic skyrmions: advances in physics and
  potential applications},\ }\href@noop {} {\bibfield  {journal} {\bibinfo
  {journal} {Nat. Rev. Mater.}\ }\textbf {\bibinfo {volume} {2}},\ \bibinfo
  {pages} {17031} (\bibinfo {year} {2017})}\BibitemShut {NoStop}%
\bibitem [{\citenamefont {Chauleau}\ \emph {et~al.}(2020)\citenamefont
  {Chauleau}, \citenamefont {Chirac}, \citenamefont {Fusil}, \citenamefont
  {Garcia}, \citenamefont {Akhtar}, \citenamefont {Tranchida}, \citenamefont
  {Thibaudeau}, \citenamefont {Gross}, \citenamefont {Blouzon}, \citenamefont
  {Finco}, \citenamefont {Bibes}, \citenamefont {Dkhil}, \citenamefont
  {Khalyavin}, \citenamefont {Manuel}, \citenamefont {Jacques}, \citenamefont
  {Jaouen},\ and\ \citenamefont {Viret}}]{Chauleau:NM}%
  \BibitemOpen
  \bibfield  {author} {\bibinfo {author} {\bibfnamefont {J.-Y.}\ \bibnamefont
  {Chauleau}}, \bibinfo {author} {\bibfnamefont {T.}~\bibnamefont {Chirac}},
  \bibinfo {author} {\bibfnamefont {S.}~\bibnamefont {Fusil}}, \bibinfo
  {author} {\bibfnamefont {V.}~\bibnamefont {Garcia}}, \bibinfo {author}
  {\bibfnamefont {W.}~\bibnamefont {Akhtar}}, \bibinfo {author} {\bibfnamefont
  {J.}~\bibnamefont {Tranchida}}, \bibinfo {author} {\bibfnamefont
  {P.}~\bibnamefont {Thibaudeau}}, \bibinfo {author} {\bibfnamefont
  {I.}~\bibnamefont {Gross}}, \bibinfo {author} {\bibfnamefont
  {C.}~\bibnamefont {Blouzon}}, \bibinfo {author} {\bibfnamefont
  {A.}~\bibnamefont {Finco}}, \bibinfo {author} {\bibfnamefont
  {M.}~\bibnamefont {Bibes}}, \bibinfo {author} {\bibfnamefont
  {B.}~\bibnamefont {Dkhil}}, \bibinfo {author} {\bibfnamefont {D.~D.}\
  \bibnamefont {Khalyavin}}, \bibinfo {author} {\bibfnamefont {P.}~\bibnamefont
  {Manuel}}, \bibinfo {author} {\bibfnamefont {V.}~\bibnamefont {Jacques}},
  \bibinfo {author} {\bibfnamefont {N.}~\bibnamefont {Jaouen}},\ and\ \bibinfo
  {author} {\bibfnamefont {M.}~\bibnamefont {Viret}},\ }\bibfield  {title}
  {\bibinfo {title} {Electric and antiferromagnetic chiral textures at
  multiferroic domain walls},\ }\href@noop {} {\bibfield  {journal} {\bibinfo
  {journal} {Nat. Mater.}\ }\textbf {\bibinfo {volume} {19}},\ \bibinfo {pages}
  {386} (\bibinfo {year} {2020})}\BibitemShut {NoStop}%
\bibitem [{\citenamefont {Guo}\ \emph {et~al.}(2021)\citenamefont {Guo},
  \citenamefont {Guo}, \citenamefont {Han}, \citenamefont {Chen}, \citenamefont
  {He}, \citenamefont {Tang}, \citenamefont {Li}, \citenamefont {Strzalka},
  \citenamefont {Ma}, \citenamefont {Yi}, \citenamefont {Wang}, \citenamefont
  {Xu}, \citenamefont {Gao}, \citenamefont {Huang}, \citenamefont {Chen},
  \citenamefont {Zhang}, \citenamefont {Lin}, \citenamefont {Nan},\ and\
  \citenamefont {Shen}}]{Guo:Science}%
  \BibitemOpen
  \bibfield  {author} {\bibinfo {author} {\bibfnamefont {M.}~\bibnamefont
  {Guo}}, \bibinfo {author} {\bibfnamefont {C.}~\bibnamefont {Guo}}, \bibinfo
  {author} {\bibfnamefont {J.}~\bibnamefont {Han}}, \bibinfo {author}
  {\bibfnamefont {S.}~\bibnamefont {Chen}}, \bibinfo {author} {\bibfnamefont
  {S.}~\bibnamefont {He}}, \bibinfo {author} {\bibfnamefont {T.}~\bibnamefont
  {Tang}}, \bibinfo {author} {\bibfnamefont {Q.}~\bibnamefont {Li}}, \bibinfo
  {author} {\bibfnamefont {J.}~\bibnamefont {Strzalka}}, \bibinfo {author}
  {\bibfnamefont {J.}~\bibnamefont {Ma}}, \bibinfo {author} {\bibfnamefont
  {D.}~\bibnamefont {Yi}}, \bibinfo {author} {\bibfnamefont {K.}~\bibnamefont
  {Wang}}, \bibinfo {author} {\bibfnamefont {B.}~\bibnamefont {Xu}}, \bibinfo
  {author} {\bibfnamefont {P.}~\bibnamefont {Gao}}, \bibinfo {author}
  {\bibfnamefont {H.}~\bibnamefont {Huang}}, \bibinfo {author} {\bibfnamefont
  {L.-Q.}\ \bibnamefont {Chen}}, \bibinfo {author} {\bibfnamefont
  {S.}~\bibnamefont {Zhang}}, \bibinfo {author} {\bibfnamefont {Y.-H.}\
  \bibnamefont {Lin}}, \bibinfo {author} {\bibfnamefont {C.-W.}\ \bibnamefont
  {Nan}},\ and\ \bibinfo {author} {\bibfnamefont {Y.}~\bibnamefont {Shen}},\
  }\bibfield  {title} {\bibinfo {title} {Toroidal polar topology in strained
  ferroelectric polymer},\ }\href@noop {} {\bibfield  {journal} {\bibinfo
  {journal} {Science}\ }\textbf {\bibinfo {volume} {371}},\ \bibinfo {pages}
  {1050} (\bibinfo {year} {2021})}\BibitemShut {NoStop}%
\bibitem [{\citenamefont {Dong}\ \emph {et~al.}(2015)\citenamefont {Dong},
  \citenamefont {Liu}, \citenamefont {Cheong},\ and\ \citenamefont
  {Ren}}]{Dong:AP}%
  \BibitemOpen
  \bibfield  {author} {\bibinfo {author} {\bibfnamefont {S.}~\bibnamefont
  {Dong}}, \bibinfo {author} {\bibfnamefont {J.-M.}\ \bibnamefont {Liu}},
  \bibinfo {author} {\bibfnamefont {S.-W.}\ \bibnamefont {Cheong}},\ and\
  \bibinfo {author} {\bibfnamefont {Z.}~\bibnamefont {Ren}},\ }\bibfield
  {title} {\bibinfo {title} {Multiferroic materials and magnetoelectric
  physics: symmetry, entanglement, excitation, and topology},\ }\href@noop {}
  {\bibfield  {journal} {\bibinfo  {journal} {Adv. Phys.}\ }\textbf {\bibinfo
  {volume} {64}},\ \bibinfo {pages} {519} (\bibinfo {year} {2015})}\BibitemShut
  {NoStop}%
\bibitem [{\citenamefont {Heron}\ \emph {et~al.}(2014)\citenamefont {Heron},
  \citenamefont {Schlom},\ and\ \citenamefont {Ramesh}}]{Heron:APR}%
  \BibitemOpen
  \bibfield  {author} {\bibinfo {author} {\bibfnamefont {J.~T.}\ \bibnamefont
  {Heron}}, \bibinfo {author} {\bibfnamefont {D.~G.}\ \bibnamefont {Schlom}},\
  and\ \bibinfo {author} {\bibfnamefont {R.}~\bibnamefont {Ramesh}},\
  }\bibfield  {title} {\bibinfo {title} {Electric field control of magnetism
  using {BiFeO$_3$}-based heterostructures},\ }\href@noop {} {\bibfield
  {journal} {\bibinfo  {journal} {Appl. Phys. Rev.}\ }\textbf {\bibinfo
  {volume} {1}},\ \bibinfo {pages} {021303} (\bibinfo {year}
  {2014})}\BibitemShut {NoStop}%
\bibitem [{\citenamefont {Wang}\ \emph
  {et~al.}(2023{\natexlab{a}})\citenamefont {Wang}, \citenamefont {Zhou},
  \citenamefont {Shen}, \citenamefont {Dong},\ and\ \citenamefont
  {Zhang}}]{Wang:PRA}%
  \BibitemOpen
  \bibfield  {author} {\bibinfo {author} {\bibfnamefont {F.}~\bibnamefont
  {Wang}}, \bibinfo {author} {\bibfnamefont {Y.}~\bibnamefont {Zhou}}, \bibinfo
  {author} {\bibfnamefont {X.}~\bibnamefont {Shen}}, \bibinfo {author}
  {\bibfnamefont {S.}~\bibnamefont {Dong}},\ and\ \bibinfo {author}
  {\bibfnamefont {J.}~\bibnamefont {Zhang}},\ }\bibfield  {title} {\bibinfo
  {title} {Magnetoelectric coupling and cross control in two-dimensional
  ferromagnets},\ }\href@noop {} {\bibfield  {journal} {\bibinfo  {journal}
  {Phys. Rev. Appl.}\ }\textbf {\bibinfo {volume} {20}},\ \bibinfo {pages}
  {064011} (\bibinfo {year} {2023}{\natexlab{a}})}\BibitemShut {NoStop}%
\bibitem [{\citenamefont {Dong}\ \emph {et~al.}(2019)\citenamefont {Dong},
  \citenamefont {Xiang},\ and\ \citenamefont {Dagotto}}]{Dong:NSR}%
  \BibitemOpen
  \bibfield  {author} {\bibinfo {author} {\bibfnamefont {S.}~\bibnamefont
  {Dong}}, \bibinfo {author} {\bibfnamefont {H.}~\bibnamefont {Xiang}},\ and\
  \bibinfo {author} {\bibfnamefont {E.}~\bibnamefont {Dagotto}},\ }\bibfield
  {title} {\bibinfo {title} {Magnetoelectricity in multiferroics: a theoretical
  perspective},\ }\href@noop {} {\bibfield  {journal} {\bibinfo  {journal}
  {Natl. Sci. Rev.}\ }\textbf {\bibinfo {volume} {6}},\ \bibinfo {pages} {629}
  (\bibinfo {year} {2019})}\BibitemShut {NoStop}%
\bibitem [{\citenamefont {Wang}\ \emph {et~al.}(2014)\citenamefont {Wang},
  \citenamefont {Hu}, \citenamefont {Ma}, \citenamefont {Zhang}, \citenamefont
  {Chen},\ and\ \citenamefont {Nan}}]{Wang:SR}%
  \BibitemOpen
  \bibfield  {author} {\bibinfo {author} {\bibfnamefont {J.~J.}\ \bibnamefont
  {Wang}}, \bibinfo {author} {\bibfnamefont {J.~M.}\ \bibnamefont {Hu}},
  \bibinfo {author} {\bibfnamefont {J.}~\bibnamefont {Ma}}, \bibinfo {author}
  {\bibfnamefont {J.}~\bibnamefont {Zhang}}, \bibinfo {author} {\bibfnamefont
  {L.-Q.}\ \bibnamefont {Chen}},\ and\ \bibinfo {author} {\bibfnamefont
  {C.-W.}\ \bibnamefont {Nan}},\ }\bibfield  {title} {\bibinfo {title} {Full
  $180^\circ$ magnetization reversal with electric fields},\ }\href@noop {}
  {\bibfield  {journal} {\bibinfo  {journal} {Sci. Rep.}\ }\textbf {\bibinfo
  {volume} {4}},\ \bibinfo {pages} {7507} (\bibinfo {year} {2014})}\BibitemShut
  {NoStop}%
\bibitem [{\citenamefont {Weng}\ \emph {et~al.}(2016)\citenamefont {Weng},
  \citenamefont {Lin}, \citenamefont {Dagotto},\ and\ \citenamefont
  {Dong}}]{Weng:PRL}%
  \BibitemOpen
  \bibfield  {author} {\bibinfo {author} {\bibfnamefont {Y.}~\bibnamefont
  {Weng}}, \bibinfo {author} {\bibfnamefont {L.}~\bibnamefont {Lin}}, \bibinfo
  {author} {\bibfnamefont {E.}~\bibnamefont {Dagotto}},\ and\ \bibinfo {author}
  {\bibfnamefont {S.}~\bibnamefont {Dong}},\ }\bibfield  {title} {\bibinfo
  {title} {Inversion of ferrimagnetic magnetization by ferroelectric switching
  via a novel magnetoelectric coupling},\ }\href@noop {} {\bibfield  {journal}
  {\bibinfo  {journal} {Phys. Rev. Lett.}\ }\textbf {\bibinfo {volume} {117}},\
  \bibinfo {pages} {037601} (\bibinfo {year} {2016})}\BibitemShut {NoStop}%
\bibitem [{\citenamefont {Lin}\ \emph {et~al.}(2017)\citenamefont {Lin},
  \citenamefont {Xu}, \citenamefont {Zhang}, \citenamefont {Zhang},
  \citenamefont {Liang},\ and\ \citenamefont {Dong}}]{Lin:PRM}%
  \BibitemOpen
  \bibfield  {author} {\bibinfo {author} {\bibfnamefont {L.~F.}\ \bibnamefont
  {Lin}}, \bibinfo {author} {\bibfnamefont {Q.~R.}\ \bibnamefont {Xu}},
  \bibinfo {author} {\bibfnamefont {Y.}~\bibnamefont {Zhang}}, \bibinfo
  {author} {\bibfnamefont {J.~J.}\ \bibnamefont {Zhang}}, \bibinfo {author}
  {\bibfnamefont {Y.~P.}\ \bibnamefont {Liang}},\ and\ \bibinfo {author}
  {\bibfnamefont {S.}~\bibnamefont {Dong}},\ }\bibfield  {title} {\bibinfo
  {title} {Ferroelectric ferrimagnetic {LiFe$_2$F$_6$}: charge ordering
  mediated magnetoelectricity},\ }\href@noop {} {\bibfield  {journal} {\bibinfo
   {journal} {Phys. Rev. Mater.}\ }\textbf {\bibinfo {volume} {1}},\ \bibinfo
  {pages} {071401(R)} (\bibinfo {year} {2017})}\BibitemShut {NoStop}%
\bibitem [{\citenamefont {Chen}\ and\ \citenamefont {Dong}(2021)}]{Chen:PRL}%
  \BibitemOpen
  \bibfield  {author} {\bibinfo {author} {\bibfnamefont {J.}~\bibnamefont
  {Chen}}\ and\ \bibinfo {author} {\bibfnamefont {S.}~\bibnamefont {Dong}},\
  }\bibfield  {title} {\bibinfo {title} {Manipulation of magnetic domain wall
  by ferroelectric switching: Dynamic magnetoelectricity at the nanoscale},\
  }\href@noop {} {\bibfield  {journal} {\bibinfo  {journal} {Phys. Rev. Lett.}\
  }\textbf {\bibinfo {volume} {126}},\ \bibinfo {pages} {117603} (\bibinfo
  {year} {2021})}\BibitemShut {NoStop}%
\bibitem [{\citenamefont {Ponet}\ \emph {et~al.}(2022)\citenamefont {Ponet},
  \citenamefont {Artyukhin}, \citenamefont {Th.Kain}, \citenamefont
  {Wettstein}, \citenamefont {Pimenov}, \citenamefont {Shuvaev}, \citenamefont
  {Wang}, \citenamefont {Cheong}, \citenamefont {Mostovoy},\ and\ \citenamefont
  {Pimenov}}]{Ponet:Nature}%
  \BibitemOpen
  \bibfield  {author} {\bibinfo {author} {\bibfnamefont {L.}~\bibnamefont
  {Ponet}}, \bibinfo {author} {\bibfnamefont {S.}~\bibnamefont {Artyukhin}},
  \bibinfo {author} {\bibnamefont {Th.Kain}}, \bibinfo {author} {\bibfnamefont
  {J.}~\bibnamefont {Wettstein}}, \bibinfo {author} {\bibfnamefont
  {A.}~\bibnamefont {Pimenov}}, \bibinfo {author} {\bibfnamefont
  {A.}~\bibnamefont {Shuvaev}}, \bibinfo {author} {\bibfnamefont
  {X.}~\bibnamefont {Wang}}, \bibinfo {author} {\bibfnamefont {S.-W.}\
  \bibnamefont {Cheong}}, \bibinfo {author} {\bibfnamefont {M.}~\bibnamefont
  {Mostovoy}},\ and\ \bibinfo {author} {\bibfnamefont {A.}~\bibnamefont
  {Pimenov}},\ }\bibfield  {title} {\bibinfo {title} {Topologically protected
  magnetoelectric switching in a multiferroic},\ }\href@noop {} {\bibfield
  {journal} {\bibinfo  {journal} {Nature}\ }\textbf {\bibinfo {volume} {607}},\
  \bibinfo {pages} {81} (\bibinfo {year} {2022})}\BibitemShut {NoStop}%
\bibitem [{\citenamefont {Liu}\ \emph {et~al.}(2022)\citenamefont {Liu},
  \citenamefont {Pi}, \citenamefont {Zhou}, \citenamefont {Liu}, \citenamefont
  {Shen}, \citenamefont {Ye}, \citenamefont {Qin}, \citenamefont {Mi},
  \citenamefont {Chen}, \citenamefont {Zhao}, \citenamefont {Zhou},
  \citenamefont {Guo}, \citenamefont {Yu}, \citenamefont {Chai}, \citenamefont
  {Weng},\ and\ \citenamefont {Long}}]{Liu:NC}%
  \BibitemOpen
  \bibfield  {author} {\bibinfo {author} {\bibfnamefont {G.}~\bibnamefont
  {Liu}}, \bibinfo {author} {\bibfnamefont {M.}~\bibnamefont {Pi}}, \bibinfo
  {author} {\bibfnamefont {L.}~\bibnamefont {Zhou}}, \bibinfo {author}
  {\bibfnamefont {Z.}~\bibnamefont {Liu}}, \bibinfo {author} {\bibfnamefont
  {X.}~\bibnamefont {Shen}}, \bibinfo {author} {\bibfnamefont {X.}~\bibnamefont
  {Ye}}, \bibinfo {author} {\bibfnamefont {S.}~\bibnamefont {Qin}}, \bibinfo
  {author} {\bibfnamefont {X.}~\bibnamefont {Mi}}, \bibinfo {author}
  {\bibfnamefont {X.}~\bibnamefont {Chen}}, \bibinfo {author} {\bibfnamefont
  {L.}~\bibnamefont {Zhao}}, \bibinfo {author} {\bibfnamefont {B.}~\bibnamefont
  {Zhou}}, \bibinfo {author} {\bibfnamefont {J.}~\bibnamefont {Guo}}, \bibinfo
  {author} {\bibfnamefont {X.}~\bibnamefont {Yu}}, \bibinfo {author}
  {\bibfnamefont {Y.}~\bibnamefont {Chai}}, \bibinfo {author} {\bibfnamefont
  {H.}~\bibnamefont {Weng}},\ and\ \bibinfo {author} {\bibfnamefont
  {Y.}~\bibnamefont {Long}},\ }\bibfield  {title} {\bibinfo {title} {Physical
  realization of topological roman surface by spin-induced ferroelectric
  polarization in cubic lattice},\ }\href@noop {} {\bibfield  {journal}
  {\bibinfo  {journal} {Nat. Commun.}\ }\textbf {\bibinfo {volume} {13}},\
  \bibinfo {pages} {2373} (\bibinfo {year} {2022})}\BibitemShut {NoStop}%
\bibitem [{\citenamefont {Wang}\ \emph
  {et~al.}(2023{\natexlab{b}})\citenamefont {Wang}, \citenamefont {Chai},\ and\
  \citenamefont {Dong}}]{Wang:PRB23}%
  \BibitemOpen
  \bibfield  {author} {\bibinfo {author} {\bibfnamefont {Z.}~\bibnamefont
  {Wang}}, \bibinfo {author} {\bibfnamefont {Y.}~\bibnamefont {Chai}},\ and\
  \bibinfo {author} {\bibfnamefont {S.}~\bibnamefont {Dong}},\ }\bibfield
  {title} {\bibinfo {title} {First-principles demonstration of roman surface
  topological multiferroicity},\ }\href@noop {} {\bibfield  {journal} {\bibinfo
   {journal} {Phys. Rev. B}\ }\textbf {\bibinfo {volume} {108}},\ \bibinfo
  {pages} {L060407} (\bibinfo {year} {2023}{\natexlab{b}})}\BibitemShut
  {NoStop}%
\bibitem [{\citenamefont {Mee}\ and\ \citenamefont {Corbett}(1964)}]{Mee:IC}%
  \BibitemOpen
  \bibfield  {author} {\bibinfo {author} {\bibfnamefont {J.~E.}\ \bibnamefont
  {Mee}}\ and\ \bibinfo {author} {\bibfnamefont {J.~D.}\ \bibnamefont
  {Corbett}},\ }\bibfield  {title} {\bibinfo {title} {Rare earth metal-metal
  halide systems. {VII}. the phases gadolinium 1.6-chloride and gadolinium
  diiodide},\ }\href@noop {} {\bibfield  {journal} {\bibinfo  {journal} {Inorg.
  Chem.}\ }\textbf {\bibinfo {volume} {4}},\ \bibinfo {pages} {88} (\bibinfo
  {year} {1964})}\BibitemShut {NoStop}%
\bibitem [{\citenamefont {Kresse}\ and\ \citenamefont
  {Furthm\"uller}(1996)}]{Kresse:PRB}%
  \BibitemOpen
  \bibfield  {author} {\bibinfo {author} {\bibfnamefont {G.}~\bibnamefont
  {Kresse}}\ and\ \bibinfo {author} {\bibfnamefont {J.}~\bibnamefont
  {Furthm\"uller}},\ }\bibfield  {title} {\bibinfo {title} {Efficient iterative
  schemes for ab initio total-energy calculations using a plane-wave basis
  set},\ }\href@noop {} {\bibfield  {journal} {\bibinfo  {journal} {Phys. Rev.
  B}\ }\textbf {\bibinfo {volume} {54}},\ \bibinfo {pages} {11169} (\bibinfo
  {year} {1996})}\BibitemShut {NoStop}%
\bibitem [{\citenamefont {Kresse}\ and\ \citenamefont
  {Furthm\"ueller}(1996)}]{Kresse:CMS}%
  \BibitemOpen
  \bibfield  {author} {\bibinfo {author} {\bibfnamefont {G.}~\bibnamefont
  {Kresse}}\ and\ \bibinfo {author} {\bibfnamefont {J.}~\bibnamefont
  {Furthm\"ueller}},\ }\bibfield  {title} {\bibinfo {title} {Efficiency of
  ab-initio total energy calculations for metals and semiconductors using a
  plane-wave basis set},\ }\href@noop {} {\bibfield  {journal} {\bibinfo
  {journal} {Comp. Mater. Sci.}\ }\textbf {\bibinfo {volume} {6}},\ \bibinfo
  {pages} {15} (\bibinfo {year} {1996})}\BibitemShut {NoStop}%
\bibitem [{\citenamefont {Blöchl}(1994)}]{BLOCHL:PRB}%
  \BibitemOpen
  \bibfield  {author} {\bibinfo {author} {\bibfnamefont {P.~E.}\ \bibnamefont
  {Blöchl}},\ }\bibfield  {title} {\bibinfo {title} {Projector augmented-wave
  method},\ }\href@noop {} {\bibfield  {journal} {\bibinfo  {journal} {Phys.
  Rev. B}\ }\textbf {\bibinfo {volume} {50}},\ \bibinfo {pages} {17953}
  (\bibinfo {year} {1994})}\BibitemShut {NoStop}%
\bibitem [{\citenamefont {Kresse}\ and\ \citenamefont
  {Joubert}(1999)}]{Kresse:PRB1}%
  \BibitemOpen
  \bibfield  {author} {\bibinfo {author} {\bibfnamefont {G.}~\bibnamefont
  {Kresse}}\ and\ \bibinfo {author} {\bibfnamefont {D.~P.}\ \bibnamefont
  {Joubert}},\ }\bibfield  {title} {\bibinfo {title} {From ultrasoft
  pseudopotentials to the projector augmented-wave method},\ }\href@noop {}
  {\bibfield  {journal} {\bibinfo  {journal} {Phys. Rev. B}\ }\textbf {\bibinfo
  {volume} {59}},\ \bibinfo {pages} {1758} (\bibinfo {year}
  {1999})}\BibitemShut {NoStop}%
\bibitem [{\citenamefont {Wang}\ and\ \citenamefont {Perdew}(1991)}]{Wang:PRB}%
  \BibitemOpen
  \bibfield  {author} {\bibinfo {author} {\bibfnamefont {Y.}~\bibnamefont
  {Wang}}\ and\ \bibinfo {author} {\bibfnamefont {J.~P.}\ \bibnamefont
  {Perdew}},\ }\bibfield  {title} {\bibinfo {title} {Correlation hole of the
  spin-polarized electron gas, with exact small-wave-vector and high-density
  scaling},\ }\href@noop {} {\bibfield  {journal} {\bibinfo  {journal} {Phys.
  Rev. B}\ }\textbf {\bibinfo {volume} {44}},\ \bibinfo {pages} {13298}
  (\bibinfo {year} {1991})}\BibitemShut {NoStop}%
\bibitem [{\citenamefont {Dudarev}\ \emph {et~al.}(1998)\citenamefont
  {Dudarev}, \citenamefont {Botton}, \citenamefont {Savrasov}, \citenamefont
  {Humphreys},\ and\ \citenamefont {Sutton}}]{Dudarev:PRB}%
  \BibitemOpen
  \bibfield  {author} {\bibinfo {author} {\bibfnamefont {S.~L.}\ \bibnamefont
  {Dudarev}}, \bibinfo {author} {\bibfnamefont {G.~A.}\ \bibnamefont {Botton}},
  \bibinfo {author} {\bibfnamefont {S.~Y.}\ \bibnamefont {Savrasov}}, \bibinfo
  {author} {\bibfnamefont {C.~J.}\ \bibnamefont {Humphreys}},\ and\ \bibinfo
  {author} {\bibfnamefont {A.~P.}\ \bibnamefont {Sutton}},\ }\bibfield  {title}
  {\bibinfo {title} {Electron-energy-loss spectra and the structural stability
  of nickel oxide: An {LSDA+U} study},\ }\href@noop {} {\bibfield  {journal}
  {\bibinfo  {journal} {Phys. Rev. B}\ }\textbf {\bibinfo {volume} {57}},\
  \bibinfo {pages} {1505} (\bibinfo {year} {1998})}\BibitemShut {NoStop}%
\bibitem [{\citenamefont {Wang}\ \emph {et~al.}(2020)\citenamefont {Wang},
  \citenamefont {Zhang}, \citenamefont {Zhang}, \citenamefont {Yuan},
  \citenamefont {Guo}, \citenamefont {Dong},\ and\ \citenamefont
  {Wang}}]{Wang:MH}%
  \BibitemOpen
  \bibfield  {author} {\bibinfo {author} {\bibfnamefont {B.}~\bibnamefont
  {Wang}}, \bibinfo {author} {\bibfnamefont {X.}~\bibnamefont {Zhang}},
  \bibinfo {author} {\bibfnamefont {Y.}~\bibnamefont {Zhang}}, \bibinfo
  {author} {\bibfnamefont {S.}~\bibnamefont {Yuan}}, \bibinfo {author}
  {\bibfnamefont {Y.}~\bibnamefont {Guo}}, \bibinfo {author} {\bibfnamefont
  {S.}~\bibnamefont {Dong}},\ and\ \bibinfo {author} {\bibfnamefont
  {J.}~\bibnamefont {Wang}},\ }\bibfield  {title} {\bibinfo {title} {Prediction
  of a two-dimensional high-{$T_C$} f-electron ferromagnetic semiconductor},\
  }\href@noop {} {\bibfield  {journal} {\bibinfo  {journal} {Mater. Horizons}\
  }\textbf {\bibinfo {volume} {7}},\ \bibinfo {pages} {1623} (\bibinfo {year}
  {2020})}\BibitemShut {NoStop}%
\bibitem [{\citenamefont {Krukau}\ \emph {et~al.}(2006)\citenamefont {Krukau},
  \citenamefont {Vydrov}, \citenamefont {Izmaylov},\ and\ \citenamefont
  {Scuseria}}]{Krukau:JCP}%
  \BibitemOpen
  \bibfield  {author} {\bibinfo {author} {\bibfnamefont {A.~V.}\ \bibnamefont
  {Krukau}}, \bibinfo {author} {\bibfnamefont {O.~A.}\ \bibnamefont {Vydrov}},
  \bibinfo {author} {\bibfnamefont {A.~F.}\ \bibnamefont {Izmaylov}},\ and\
  \bibinfo {author} {\bibfnamefont {G.~E.}\ \bibnamefont {Scuseria}},\
  }\bibfield  {title} {\bibinfo {title} {Influence of the exchange screening
  parameter on the performance of screened hybrid functionals},\ }\href@noop {}
  {\bibfield  {journal} {\bibinfo  {journal} {J. Chem. Phys.}\ }\textbf
  {\bibinfo {volume} {125}},\ \bibinfo {pages} {224106} (\bibinfo {year}
  {2006})}\BibitemShut {NoStop}%
\bibitem [{\citenamefont {Togo}\ and\ \citenamefont {Tanaka}(2015)}]{Togo:SM}%
  \BibitemOpen
  \bibfield  {author} {\bibinfo {author} {\bibfnamefont {A.}~\bibnamefont
  {Togo}}\ and\ \bibinfo {author} {\bibfnamefont {I.}~\bibnamefont {Tanaka}},\
  }\bibfield  {title} {\bibinfo {title} {First principles phonon calculations
  in materials science},\ }\href@noop {} {\bibfield  {journal} {\bibinfo
  {journal} {Scripta Mater.}\ }\textbf {\bibinfo {volume} {108}},\ \bibinfo
  {pages} {1} (\bibinfo {year} {2015})}\BibitemShut {NoStop}%
\bibitem [{\citenamefont {King-Smith}\ and\ \citenamefont
  {Vanderbilt}(1993)}]{KINGSMITH:PRB}%
  \BibitemOpen
  \bibfield  {author} {\bibinfo {author} {\bibfnamefont {R.~D.}\ \bibnamefont
  {King-Smith}}\ and\ \bibinfo {author} {\bibfnamefont {D.}~\bibnamefont
  {Vanderbilt}},\ }\bibfield  {title} {\bibinfo {title} {Theory of polarization
  of crystalline solids},\ }\href@noop {} {\bibfield  {journal} {\bibinfo
  {journal} {Phys. Rev. B}\ }\textbf {\bibinfo {volume} {47}},\ \bibinfo
  {pages} {1651} (\bibinfo {year} {1993})}\BibitemShut {NoStop}%
\bibitem [{\citenamefont {Xiao}\ \emph {et~al.}(2012)\citenamefont {Xiao},
  \citenamefont {Liu}, \citenamefont {Feng}, \citenamefont {Xu},\ and\
  \citenamefont {Yao}}]{Xiao:Prl}%
  \BibitemOpen
  \bibfield  {author} {\bibinfo {author} {\bibfnamefont {D.}~\bibnamefont
  {Xiao}}, \bibinfo {author} {\bibfnamefont {G.-B.}\ \bibnamefont {Liu}},
  \bibinfo {author} {\bibfnamefont {W.}~\bibnamefont {Feng}}, \bibinfo {author}
  {\bibfnamefont {X.}~\bibnamefont {Xu}},\ and\ \bibinfo {author}
  {\bibfnamefont {W.}~\bibnamefont {Yao}},\ }\bibfield  {title} {\bibinfo
  {title} {Coupled spin and valley physics in monolayers of {MoS$_2$} and other
  {Group-VI} dichalcogenides},\ }\href@noop {} {\bibfield  {journal} {\bibinfo
  {journal} {Phys. Rev. Lett.}\ }\textbf {\bibinfo {volume} {108}},\ \bibinfo
  {pages} {196802} (\bibinfo {year} {2012})}\BibitemShut {NoStop}%
\bibitem [{sm()}]{sm}%
  \BibitemOpen
  \href@noop {} {}\bibinfo {note} {See Supplymenntal Material for electronic
  structure, magnetic ground state, magnetic symmetry analysis, more results of
  micromagnetic simulations of GdI$_2$ monolayer, and derivation of the
  magnetoelectric tensors for 2D point groups, including
  Ref.~\cite{Matsumoto:JPSJ}}\BibitemShut {NoStop}%
\bibitem [{\citenamefont {Kasten}\ \emph {et~al.}(1984)\citenamefont {Kasten},
  \citenamefont {Mueller},\ and\ \citenamefont {Schienle}}]{KASTEN:SSC}%
  \BibitemOpen
  \bibfield  {author} {\bibinfo {author} {\bibfnamefont {A.}~\bibnamefont
  {Kasten}}, \bibinfo {author} {\bibfnamefont {P.~H.~J.}\ \bibnamefont
  {Mueller}},\ and\ \bibinfo {author} {\bibfnamefont {M.}~\bibnamefont
  {Schienle}},\ }\bibfield  {title} {\bibinfo {title} {Magnetic ordering in
  {GdI$_2$}},\ }\href@noop {} {\bibfield  {journal} {\bibinfo  {journal} {Solid
  State Commun.}\ }\textbf {\bibinfo {volume} {51}},\ \bibinfo {pages} {919}
  (\bibinfo {year} {1984})}\BibitemShut {NoStop}%
\bibitem [{\citenamefont {Murakawa}\ \emph {et~al.}(2010)\citenamefont
  {Murakawa}, \citenamefont {Onose}, \citenamefont {Miyahara}, \citenamefont
  {Furukawa},\ and\ \citenamefont {Tokura}}]{Murakawa:PRL}%
  \BibitemOpen
  \bibfield  {author} {\bibinfo {author} {\bibfnamefont {H.}~\bibnamefont
  {Murakawa}}, \bibinfo {author} {\bibfnamefont {Y.}~\bibnamefont {Onose}},
  \bibinfo {author} {\bibfnamefont {S.}~\bibnamefont {Miyahara}}, \bibinfo
  {author} {\bibfnamefont {N.}~\bibnamefont {Furukawa}},\ and\ \bibinfo
  {author} {\bibfnamefont {Y.}~\bibnamefont {Tokura}},\ }\bibfield  {title}
  {\bibinfo {title} {Ferroelectricity induced by spin-dependent metal-ligand
  hybridization in {Ba$_2$CoGe$_2$O$_7$}},\ }\href@noop {} {\bibfield
  {journal} {\bibinfo  {journal} {Phys. Rev. Lett.}\ }\textbf {\bibinfo
  {volume} {105}},\ \bibinfo {pages} {137202} (\bibinfo {year}
  {2010})}\BibitemShut {NoStop}%
\bibitem [{\citenamefont {Zhang}\ \emph {et~al.}(2022)\citenamefont {Zhang},
  \citenamefont {Zhou}, \citenamefont {Wang}, \citenamefont {Shen},
  \citenamefont {Wang},\ and\ \citenamefont {Lu}}]{Zhang:PRL}%
  \BibitemOpen
  \bibfield  {author} {\bibinfo {author} {\bibfnamefont {J.}~\bibnamefont
  {Zhang}}, \bibinfo {author} {\bibfnamefont {Y.}~\bibnamefont {Zhou}},
  \bibinfo {author} {\bibfnamefont {F.}~\bibnamefont {Wang}}, \bibinfo {author}
  {\bibfnamefont {X.}~\bibnamefont {Shen}}, \bibinfo {author} {\bibfnamefont
  {J.}~\bibnamefont {Wang}},\ and\ \bibinfo {author} {\bibfnamefont
  {X.}~\bibnamefont {Lu}},\ }\bibfield  {title} {\bibinfo {title} {Coexistence
  and coupling of spin-induced ferroelectricity and ferromagnetism in
  perovskites},\ }\href@noop {} {\bibfield  {journal} {\bibinfo  {journal}
  {Phys. Rev. Lett.}\ }\textbf {\bibinfo {volume} {129}},\ \bibinfo {pages}
  {117603} (\bibinfo {year} {2022})}\BibitemShut {NoStop}%
\bibitem [{\citenamefont {Arima}(2007)}]{Arima:JPSJ}%
  \BibitemOpen
  \bibfield  {author} {\bibinfo {author} {\bibfnamefont {T.-H.}\ \bibnamefont
  {Arima}},\ }\bibfield  {title} {\bibinfo {title} {Ferroelectricity induced by
  proper-screw type magnetic order},\ }\href@noop {} {\bibfield  {journal}
  {\bibinfo  {journal} {J. Phys. Soc. Jpn.}\ }\textbf {\bibinfo {volume}
  {76}},\ \bibinfo {pages} {073702} (\bibinfo {year} {2007})}\BibitemShut
  {NoStop}%
\bibitem [{\citenamefont {Jia}\ \emph {et~al.}(2006)\citenamefont {Jia},
  \citenamefont {Onoda}, \citenamefont {Nagaosa},\ and\ \citenamefont
  {Han}}]{Jia:PRB}%
  \BibitemOpen
  \bibfield  {author} {\bibinfo {author} {\bibfnamefont {C.}~\bibnamefont
  {Jia}}, \bibinfo {author} {\bibfnamefont {S.}~\bibnamefont {Onoda}}, \bibinfo
  {author} {\bibfnamefont {N.}~\bibnamefont {Nagaosa}},\ and\ \bibinfo {author}
  {\bibfnamefont {J.~H.}\ \bibnamefont {Han}},\ }\bibfield  {title} {\bibinfo
  {title} {Bond electronic polarization induced by spin},\ }\href@noop {}
  {\bibfield  {journal} {\bibinfo  {journal} {Phys. Rev. B}\ }\textbf {\bibinfo
  {volume} {74}},\ \bibinfo {pages} {224444} (\bibinfo {year}
  {2006})}\BibitemShut {NoStop}%
\bibitem [{\citenamefont {Matsumoto}\ \emph {et~al.}(2017)\citenamefont
  {Matsumoto}, \citenamefont {Chimata},\ and\ \citenamefont
  {Koga}}]{Matsumoto:JPSJ}%
  \BibitemOpen
  \bibfield  {author} {\bibinfo {author} {\bibfnamefont {M.}~\bibnamefont
  {Matsumoto}}, \bibinfo {author} {\bibfnamefont {K.}~\bibnamefont {Chimata}},\
  and\ \bibinfo {author} {\bibfnamefont {M.}~\bibnamefont {Koga}},\ }\bibfield
  {title} {\bibinfo {title} {Symmetry analysis of spin-dependent electric
  dipole and its application to magnetoelectric effects},\ }\href@noop {}
  {\bibfield  {journal} {\bibinfo  {journal} {J. Phys. Soc. Jpn.}\ }\textbf
  {\bibinfo {volume} {86}},\ \bibinfo {pages} {034704} (\bibinfo {year}
  {2017})}\BibitemShut {NoStop}%
\end{thebibliography}%
\end{document}